\documentclass[11pt]{article}
\usepackage{coling2018}
\usepackage{times}
\usepackage{url}
\usepackage{latexsym}

\usepackage{url}
\usepackage{helvet}
\usepackage{courier}
\usepackage{graphicx}

\usepackage{adjustbox}
\usepackage{amsmath,amssymb,amsthm}
\usepackage{mathabx}
\usepackage{multirow}
\usepackage{url}
\usepackage{hhline}
\usepackage{enumerate}
\usepackage[font=small,labelfont=bf]{subcaption}
\usepackage{pgfplots}
\usepackage{xcolor}
\usetikzlibrary{arrows,intersections}

\def\figlabel#1{\label{fig:#1}\label{p:#1}}

\newcommand*{\affaddr}[1]{#1} 
\newcommand*{\affmark}[1][*]{\textsuperscript{#1}}

\title{Replicated Siamese LSTM in Ticketing System for\\ Similarity Learning and Retrieval in Asymmetric Texts}

\author{Pankaj Gupta\affmark[1,2]\\\And 
  Bernt Andrassy\affmark[1]\\
\affaddr{\affmark[1]Corporate Technology, Machine-Intelligence (MIC-DE), Siemens AG  Munich, Germany}\\
  \affaddr{\affmark[2]CIS, University of Munich (LMU) Munich, Germany} \\
  {\tt pankaj.gupta@siemens.com | bernt.andrassy@siemens.com}\\
  {\tt pankaj.gupta@campus.lmu.de |  inquiries@cis.lmu.de}\\\And
  Hinrich Sch\"{u}tze\affmark[2]\\
  }

\date{}

\begin{document}
\maketitle

\begin{abstract}
The goal of our industrial ticketing system is to retrieve a relevant solution for an input query, 
by matching with historical tickets stored in knowledge base. 
A query is comprised of subject and description, while a historical ticket consists of subject, description and solution.  
To retrieve a relevant solution, we use textual similarity paradigm to learn similarity in the query and historical tickets. 
The task is challenging due to significant term mismatch in the query and ticket pairs of asymmetric lengths, where subject is a short text 
but description and solution are multi-sentence texts. 
 We present a novel Replicated Siamese LSTM model to learn similarity in asymmetric text pairs, that 
gives 22\% and 7\% gain (Accuracy@10) for retrieval task, respectively over unsupervised and supervised baselines.
We also show that the topic and distributed semantic features for short and long texts improved both similarity learning and retrieval.
\end{abstract}

%
%
\blfootnote{
    %
    %
    %
    %
    \hspace{-0.65cm}  
     This work is licenced under a Creative Commons 
    Attribution 4.0 International Licence.
    Licence details:
    \url{http://creativecommons.org/licenses/by/4.0/}
    %
    %
}

\section{Introduction}
Semantic Textual Similarity (STS) is the task to find out if 
the text pairs mean the same thing. The important tasks in 
Natural Language Processing (NLP),  such as Information Retrieval (IR) 
and text understanding may be improved by modeling the underlying 
semantic similarity between texts. 
  
With recent progress in deep learning, the STS task has gained success using LSTM \cite{Mueller:81} and CNN \cite{Yin:81} based architectures; 
however, these approaches model the underlying semantic similarity 
between example pairs,
 each with a single sentence or phrase with term overlaps. 
In the domain of question retrieval \cite{LiCai:82,KaiZhang:82}, users retrieve   
historical questions which precisely match their
questions (single sentence) semantically equivalent or relevant. 
However, we investigate similarity learning between texts of asymmetric lengths, such as short (phrase) Vs longer (paragraph/documents) with significant term mismatch. 
The application of textual understanding in retrieval becomes more challenging when the relevant document-sized retrievals are stylistically distinct with the input short texts.    
Learning a similarity metric has gained 
much research interest, however due to limited availability of labeled data 
and complex structures in variable length sentences, the STS task  becomes a hard problem. 
The performance of IR system is sub-optimal due to significant term mismatch in similar texts \cite{Zhao:81}, limited annotated data and complex structures in variable length sentences. 
We address the challenges in a real-world industrial application.

Our ticketing system (Figure~\ref{fig:architecture}(a)) consists of a query and historical tickets (Table \ref{examplemotivation}). 
A query (reporting issue, $q$) has 2 components: {\it subject} (SUB) and {\it description} (DESC), while a historical ticket ($\mathfrak{t}$) stored in the knowledge base (KB)
has 3 components: SUB, DESC and {\it solution} (SOL).  A SUB is a short text, but DESC and SOL consist of multiple sentences. 
Table \ref{examplemotivation} shows that SUB $\in$ $q$ and SUB $\in$ $\mathfrak{t}$ are semantically similar and
 few terms in SUB $\in$ $q$ overlap with DESC $\in$ $\mathfrak{t}$. 
However, the expected SOL $\in$ $\mathfrak{t}$ is distinct from both SUB and DESC $\in$ $q$. 
The goal is to retrieve an optimal action (i.e. SOL from $\mathfrak{t}$) for the input $q$. 

To improve retrieval for an input $q$, 
we adapt the Siamese LSTM \cite{Mueller:81} for similarity learning in asymmetric text pairs, using the available information in $q$ and $\mathfrak{t}$. 
For instance, we compute {\it multi-level} similarity between (SUB $\in$ $q$, SUB $\in$ $\mathfrak{t}$) and  (DESC $\in$ $q$, DESC $\in$ $\mathfrak{t}$). 
However, observe in Table \ref{examplemotivation} that the {\it cross-level} similarities 
such as between (SUB $\in$ $q$, DESC $\in$ $\mathfrak{t}$), (DESC $\in$ $q$, SUB $\in$ $\mathfrak{t}$) 
or (SUB $\in$ $q$, SOL $\in$ $\mathfrak{t}$), etc. can supplement IR performance. See Figure~\ref{fig:architecture}(b).

\begin{table}[t]
\centering
\def\arraystretch{1.2}
\resizebox{0.96\textwidth}{!}{
\begin{tabular}{l}
\multicolumn{1}{c}{\underline{\bf QUERY} $\bf (q)$}\\
 {\bf SUB:} 
GT Trip - Low Frequency Pulsations\\
{\bf DESC:} 
GT Tripped due to a sudden  increase in Low Frequency
 Pulsations. The machine has been restarted and is now\\
 operating normally. Alarm received was: GT XXX Low Frequency Pulsation. \\ \hline
\multicolumn{1}{c}{\underline{\bf HISTORICAL TICKET} ${\bf (\mathfrak{t})}$}\\
{\bf SUB:} 
Narrow Frequency Pulsations\\ 
{\bf DESC:} 
Low and Narrow frequency pulsations were detected. The peak value for the Low Frequency Pulsations is \#\# mbar.\\
{\bf SOL:} 
XXXX combustion support is currently working on the issue.
 The 
 action 
 is that the machine 
 should  not run 
until resolved.
\end{tabular}}
\caption{Example of a Query and Historical Ticket}
\label{examplemotivation}
\end{table}

\begin{figure*}[t] 
\centering
{
\includegraphics[scale=0.7]{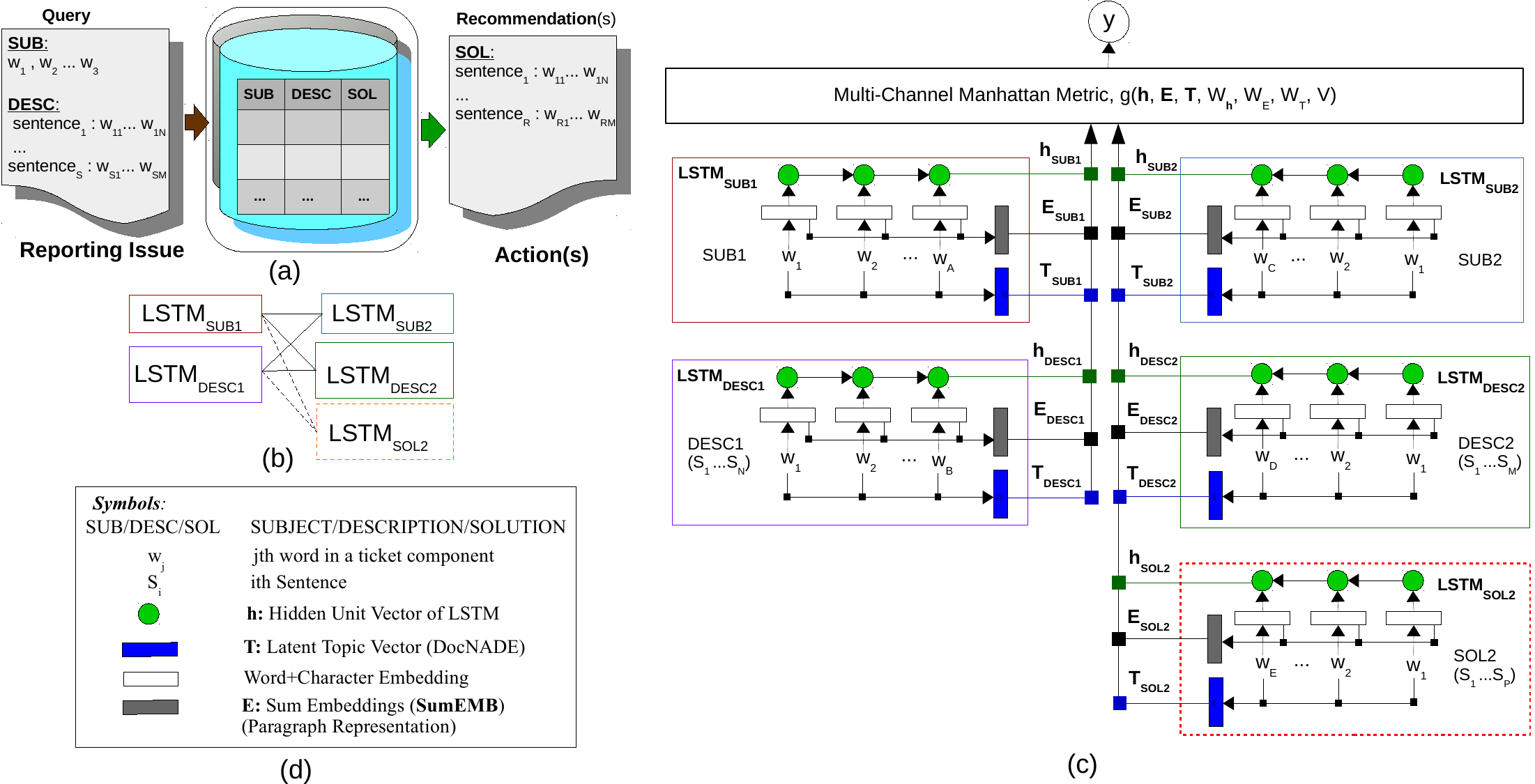}}
\caption{(a): Intelligent Ticketing System (ITS) (b): High-level illustration of Siamese LSTM 
for cross-level pairwise similarity. (c): Replicated Siamese with multi-channel (SumEMB, LSTM and topic vectors) and multi-level (SUB, DESC and/or SOL) inputs in the objective function, $g$. 
$y$: similarity score. 
The dotted lines indicate ITS output. (d): Symbols used.} 
\figlabel{architecture}
\end{figure*}

The \textit{contributions} of this paper are as follows: 
{(1)} Propose a novel architecture (Replicated Siamese LSTM) for similarity learning in asymmetric texts via multi-and-cross-level semantics 
{(2)} Investigate distributed and neural topic semantics for similarity learning via multiple channels 
{(3)} Demonstrate a gain of 22\% and 7\% in Accuracy@10 for retrieval, respectively over unsupervised and supervised baselines 
in the industrial application of a ticketing system. 

\section{Methodology}
Siamese networks \cite{Chopra:81} are
dual-branch networks with tied weights 
and an objective function.  The aim of training is 
to learn text pair representations 
to form a highly structured space where they reflect complex semantic  relationships. 
Figure~\ref{fig:architecture} shows the proposed Replicated Siamese neural network 
architecture 
 such that {\small(LSTM\textsubscript{SUB1}+LSTM\textsubscript{DESC1}) = (LSTM\textsubscript{SUB2}+LSTM\textsubscript{DESC2}+LSTM\textsubscript{SOL2})}, 
 to learn similarities in asymmetric texts, where a query {\small(SUB1+DESC1)} is stylistically 
 distinct from a historical ticket {\small(SUB2+DESC2+SOL2)}. 

Note, the {\it query components are suffixed by ``1''  and historical ticket components by ``2''} in context of the following work for pairwise comparisons.

\begin{figure*}[t]
\centering{
\begin{equation}\label{simmetric}
\small
\begin{aligned}
\begin{split}
g(& h, E, T, W_{h}, W_{E}, W_{T},   V)  = \\ 
\exp \Big( - &\sum_{p \ \in \ \{SUB1, DESC1\}}  \sum_{q \ \in \ \{SUB2, DESC2, SOL2\}} V_{\{p,q\}} \  \big( W_{h} \Vert h_{p} - h_{q} \Vert_{1} + W_{E} \Vert E_{p}  - E_{q} \Vert_{1} + W_{T} \Vert  T_{p} - T_{q} \Vert_{1} \big) \Big)
\end{split}
\end{aligned}
\end{equation}
}
\caption{Multi-Channel Manhattan Metric}
\label{Multi-Channel Manhattan Metric}
\end{figure*}

\subsection{Replicated, Multi-and-Cross-Level, Multi-Channel Siamese LSTM}\label{replicatedSiamese}
Manhattan LSTM \cite{Mueller:81} learns similarity in text pairs,
each with a single sentence;  however, we advance the similarity learning 
task in asymmetric texts pairs consisting of one or more sentences, 
where similarity is computed between different-sized subject and description or solution texts. 
As the backbone of our work, we compute similarity scores 
 to learn a highly structured space via LSTM \cite{Sepp:81} for representation of each pair of the query (SUB1 and DESC1) or historical ticket (SUB2, DESC2 and SOL2) components,  
which includes multi-level 
 ({\small SUB1-SUB2}, {\small DESC1-DESC2}) and  cross-level ({\small SUB1-DESC2},
 {\small SUB1-SOL2}, etc.) asymmetric textual similarities, Figure~\ref{fig:architecture}(b) and (c). 
To accumulate the semantics of variable-length sentences ($x_{1},...,x_{T}$), recurrent neural networks (RNNs) \cite{Thang:82,Gupta:90,guptapatentTFMTRNN}, especially the LSTMs \cite{Sepp:81} have been successful. 
 
 LSTMs are superior in learning long range dependencies through 
 their memory cells. 
 Like the standard RNN \cite{Mikolov:82,Gupta:89,Thang:81}, LSTM sequentially updates
a hidden-state representation, but it introduces a
memory state $c_{t}$ and three gates that control the
flow of information through the time steps. An
output gate $o_{t}$ determines how much of $c_{t}$ should
be exposed to the next node. An input gate $i_{t}$ controls
how much the input $x_{t}$ be stored in memory, while
the forget gate $f_{t}$ determines what should be forgotten from memory. The dynamics: 
\begin{equation}\label{LSTM}
\begin{aligned}
 i_{t} = \sigma(W_{i}x_{t} + U_{i}h_{t-1}) \\
f_{t} =  \sigma(W_{f}x_{t} + U_{f}h_{t-1}) \\
o_{t} =  \sigma(W_{o}x_{t} + U_{o}h_{t-1})  \\
\tilde{c}_{t} = tanh(W_{c}x_{t} + U_{c}h_{t-1}) \\
c_{t} = i_{t} \odot \tilde{c}_{t} + f_{t} \odot c_{t-1} \\
h_{t} = o_{t} \odot tanh(c_{t}) 
\end{aligned}
\end{equation}

where $\sigma(x)=\frac{1}{1+ e^{-x}}$ and $\tanh(x)=\frac{e^{x} - e^{-x}}{e^{x} + e^{-x}}$. 
The proposed architecture, Figure~\ref{fig:architecture}(c) is composed of multiple 
 uni-directional LSTMs 
 each for subject, description and solution within the Siamese framework, where the weights 
 at over levels are shared between the left and right branch of the network. Therefore, the name {\it replicated}. 

Each LSTM learns a mapping from space of variable length sequences, including asymmetric texts, to a hidden-state
 vector, $h$. Each sentence ($w_{1},...w_{T}$) is passed to LSTM, which updates hidden state via eq \ref{LSTM}.
A final encoded representation (e.g. h\textsubscript{SUB1}, h\textsubscript{SUB2} in Figure \ref{fig:architecture}(c)) is obtained for each query or ticket component.   
A single LSTM is run over DESC and SOL components, consisting of one or more sentences. 
Therefore, the name {\it multi-level} Siamese. 

The representations across the text components (SUB DESC or SOL) are learned in order to maximize the similarity and retrieval for a query with the historical tickets.  
Therefore, the name {\it cross-level} Siamese. 
 
The sum-average strategy over word embedding \cite{Mikolov:82} for short and longer texts has demonstrated a strong baseline for text classification 
 \cite{Jouli:81} and pairwise similarity learning \cite{Wieting:81}. 
This simple baseline to represent sentences as bag of words (BoW) inspires us to use the BoW for each query or historical ticket component, for instance $E_{SUB1}$. We refer the approach as {\it SumEMB} in the context of this paper. 

We supplement the similarity metric ($g$) with {\it SumEMB} ($E$), latent topic ($T$) (section \ref{DocNADE}) and hidden vectors ($h$) of LSTM for each  text component from both the Siamese branches.  
Therefore, the name {\it multi-channel} Siamese. \\


\begin{minipage}[t]{\textwidth}
  \begin{minipage}[b]{0.4\textwidth}
    \centering
\includegraphics[scale=0.65]{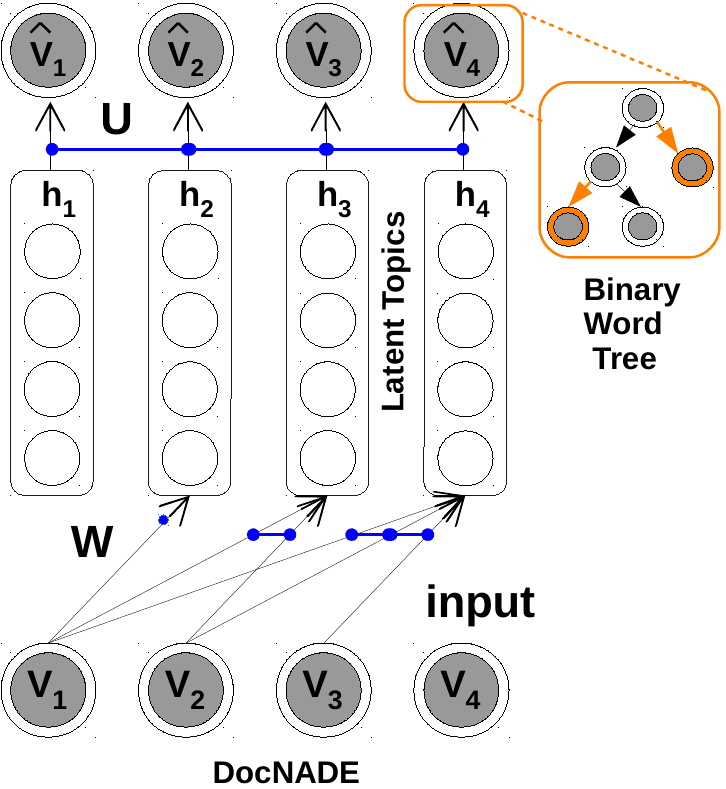}
\captionof{figure}{DocNADE: Neural Auto-regressive Topic Model}
\label{DocNADEsuppl}
  \end{minipage}
  \qquad 
  \begin{minipage}[b]{0.5\textwidth}
   \centering
\def\arraystretch{1.1}
\resizebox{0.75\textwidth}{!}{
\begin{tabular}{c|c||c}
{\it Parameter} & {\it Search} & {\it Optimal} \\ \hline
   $E$       &    [350]    &    350     \\
   $T$     &   [20, 50, 100]     &  100       \\
    $h$     &    [50, 100]    &     50    \\
     $W_{h}$    &     [0.6, 0.7, 0.8]   &   0.7 \\    
$W_{E}$    &     [0.3, 0.2, 0.1]   &   0.1 \\    
$W_{T}$    &     [0.3, 0.2, 0.1]   &   0.2 \\    
$V_{SUB1-SUB2}$    &     [0.3, 0.4   &   0.3 \\    
$V_{DESC1-DESC2}$    &     [0.3, 0.4]   &   0.3 \\   
$V_{SUB1-DESC2}$    &     [0.10, 0.15, 0.20]   &   0.20 \\   
$V_{SUB1-SOL2}$    &     [0.10, 0.15, 0.20]   &   0.10 \\   
$V_{DESC1-SOL2}$    &     [0.10, 0.15, 0.20]   &   0.10  \\ \hline
\end{tabular}}
\captionof{table}{Hyperparameters in the Replicated Siamese LSTM (experiment \#No:22)}
\label{hyperparameterssupp}\hfill
    \end{minipage}
\end{minipage}

\subsection{Neural Auto-Regressive Topic Model}\label{DocNADE}
Topic models such as Latent Dirichlet allocation (LDA) \cite{Blei:82} and Replicated Softmax (RSM) \cite{Ruslan:82,GuptaRNNRSM:82} have been popular in learning meaningful representations of unlabeled text documents. 
Recently, a new type of topic model called the Document Neural Autoregressive Distribution Estimator (DocNADE)  \cite{Larochelle:82,Larochelle:84,GuptaiDocNADE:82} was proposed and demonstrated the state-of-
the-art performance for text document modeling. 
DocNADE models are advanced variants of Restricted Boltzmann Machine \cite{hin:82,Rus:82,Gupta:86,Gupta:91}, and have shown to outperform LDA and RSM in terms of both log-likelihood of the data and document retrieval. 
In addition, the training complexity of a DocNADE model scales logarithmically with vocabulary size, instead linear as in RSM. 
The features are important for an industrial task along with quality performance. Therefore, we adopt DocNADE model for learning latent representations of tickets and retrieval in unsupervised fashion.  
See \newcite{Larochelle:82} and \newcite{GuptaiDocNADE:82} for further details, and Figure \ref{DocNADEsuppl} for the DocNADE architecture, 
where we extract the last hidden topic layer ($h_4$) to compute document representation. 

\subsection{Multi-Channel Manhattan Metric}
\newcite{Chopra:81} indicated that using $l_{2}$ instead of $l_{1}$ norm 
in similarity metric can lead to undesirable plateaus. 
\newcite{Mueller:81}  showed stable and improved results using Manhattan distance over cosine similarity. 
 
\newcite{Mueller:81} used a Manhattan metric ($l_{1}$-norm) for similarity learning in single sentence pairs. 
However, we adapt the similarity metric for 2-tuple (SUB1, DESC1) vs 3-tuple (SUB2, DESC2 and SOL2) pairs, where 
the error signals 
 are back-propagated in the multiple levels and channels during training to force the Siamese network to entirely capture the semantic differences across the query and historical tickets components. 
The similarity metric, $g$ $\in$ [0,1] is given in eq \ref{simmetric}, where $||\cdot||$ is  $l_{1}$ norm. $W_h$, $W_E$ and $W_T$ are the three channels weights for $h$, $E$ and and $T$, respectively. 
The weights ($V$) are the multi-level weights between the ticket component pairs.  Observe that a single weight is being used in the ordered ticket component pairs, for instance $V_{SUB1-DESC2}$ is same as $V_{DESC2-SUB1}$.

\begin{table}[htp]
    \begin{subtable}{.5\linewidth}
      \centering
\def\arraystretch{1.25}
      \resizebox{0.95\textwidth}{!}{
       \begin{tabular}{r|cc||cc}
        \multirow{1}{*}{\it Held-out}      & \multicolumn{4}{c}{{\it Perplexity} (100 topics)} \\ \cline{2-5}
       \multirow{1}{*}{\it Ticket}     & \multicolumn{2}{c||}{$M1$: SUB+DESC} & \multicolumn{2}{c}{$M2$: SUB+DESC+SOL} \\ \cline{2-5}
   \multirow{1}{*}{\it Component}        & LDA          & DocNADE        & LDA          & DocNADE    \\ \hline
DESC                  &       380       &         {\bf 362}           &       565       &         {\bf 351}     \\
SUB+DESC          &        480      &       {\bf 308}           &       515      &        {\bf 289}       \\
SUB+DESC+SOL &     553         &       {\bf 404 }        &       541        &     {\bf 322}   
\end{tabular}}
 \caption{}\label{pplLDAvsDocNADE}
    \end{subtable}%
    \begin{subtable}{.5\linewidth}
      \centering
\def\arraystretch{1.25}
        \resizebox{0.95\textwidth}{!}{
        \begin{tabular}{r|cc|cc}
        \multirow{1}{*}{Query}      & \multicolumn{4}{c}{Perplexity (100 topics)} \\ \cline{2-5}
     \multirow{1}{*}{Component}  & \multicolumn{2}{c|}{DocNADE:${\bf M1}$} & \multicolumn{2}{c}{DocNADE:${\bf M2}$} \\ \cline{2-5}
        & \multicolumn{1}{c}{$|Q|_{L}$}  & \multicolumn{1}{c|}{$|Q|_{U}$} & \multicolumn{1}{c}{$|Q|_{L}$}  & \multicolumn{1}{c}{$|Q|_{U}$} \\ \hline 
DESC1                            &         192                                 &           177                                                      &       \underline{132}                   &     \underline{118}             \\
SUB1+DESC1                  &    164                                       &   140                                                     &    \underline{130}                  &                \underline{118} 
\end{tabular}}
\caption{}\label{ppl}
    \end{subtable}
\caption{{\bf (a)} Perplexity by DocNADE and LDA trained with $M1$: SUB+DESC or $M2$: SUB+DESC+SOL on all tickets and evaluated on 50 held-out tickets with their respective components or their combination.
Observe that when DocNADE is trained with SUB+DESC+SOL, it performs better when training with SUB+DESC+SOL and outperforms LDA.
 {\bf (b)} Perplexity by DocNADE: $M1$ trained on SUB+ DESC and $M2$ on SUB+DESC+SOL of the historical tickets.}
\end{table}

\def\arraystretch{1.2}
\begin{table*}[t]
\centering
\small
\resizebox{\textwidth}{!}{
\begin{tabular}{l|l}
\multicolumn{1}{c|}{\textbf{Model}} & \multicolumn{1}{c}{\textbf{Model Configuration}} \\ \hline
{\it T} ($X1$--$X2$)           &    Compute Similarity using topic vector ({\it T}) pairs of a query ($X1$) and historical ticket ($X2$) components\\
{\it E} ($X1$--$X2$)           &   Compute Similarity  using embedding vector ({\it E}) pairs of a query ($X1$) and historical ticket ($X2$) components\\
$X+Y+Z$                         & Merge text components (SUB, DESC or SOL), representing a single document\\
{\it T} ($X1+Y1$--$X2+Y2+Z2$)      &    Compute Similarity using topic vector ({\it T}) pairs of a query ($X1+Y1$) and historical ticket ($X2+Y2+Z2$) components\\
S-LSTM ($X1$--$X2$)           &  Compute Similarity using Standard Siamese LSTM on a query ($X1$) and historical ticket ($X2$) components\\
ML ($X1$--$X2$, $Y1$--$Y2$)    &    Multi-level Replicated Siamese LSTM. Compute similarity in ($X1$--$X2$)  and ($Y1$--$Y2$) components of a query and historical ticket\\ 
CL (X, Y, Z)      &    Cross-level Replicated Siamese LSTM.  Compute similarity in ($X1$--$Y2$), ($X1$--$Z2$), ($Y1$--$X2$) and ($Y1$--$Z2$) pairs
\end{tabular}}
\caption{Different model configurations for the experimental setups and evaluations. See Figure~\ref{fig:architecture}(c) for LSTM configurations.}
\label{modelsettings}
\end{table*}

\def\arraystretch{1.25}
\begin{table*}[htp]
\centering
\small
\resizebox{\textwidth}{!}{
\begin{tabular}{c|r|ccc|ccccccccc}
 \multicolumn{1}{c|}{\textbf{\#No}} & \multicolumn{1}{c|}{\textbf{Model (Query-Historical Ticket)}} & \multicolumn{3}{c}{\textbf{Similarity Task}} & \multicolumn{9}{|c}{\textbf{Retrieval Task}} \\ \cline{3-14}
& & \multicolumn{1}{c}{\textit{r}} & \multicolumn{1}{c}{$\rho$} & \multicolumn{1}{c|}{MSE} & \multicolumn{1}{c}{MAP@1} & \multicolumn{1}{c}{MAP@5} & \multicolumn{1}{c}{MAP@10} & \multicolumn{1}{c}{MRR@1} & \multicolumn{1}{c}{MRR@5} & \multicolumn{1}{c}{MRR@10} & \multicolumn{1}{c}{Acc@1} & \multicolumn{1}{c}{Acc@5} & \multicolumn{1}{c}{Acc@10}\\ \hline
1 & \multicolumn{1}{l|}{T (SUB1--SUB2) ({\bf unsupervised baseline})}          & 0.388  &   0.330  &  5.122    &  0.08 & 0.08  &   0.07
  &  1.00    &  0.28  &  0.10  &  0.04  &  0.19  &  0.30 \\ 
2 &  \multicolumn{1}{l|}{T (SUB1--DESC2)}          & 0.347  &   0.312  &  3.882    &  0.09 & 0.07  &   0.07
  &  0.00    &  0.05  &  0.08  &  0.04  &  0.13  &  0.21 \\ 
3 & \multicolumn{1}{l|}{T (DESC1--SUB2)}          & 0.321  &   0.287  &  3.763    &  0.08 & 0.09  &   0.09
  &  0.00    &  0.05  &  0.11  &  0.03  &  0.20  &  0.31 \\
4 & \multicolumn{1}{l|}{T (DESC1--DESC2)}          & 0.402  &   0.350  &  3.596    &  0.08 & 0.08  &   0.08
  &  0.00    &  0.04  &  0.10  &  0.03  &  0.19  &  0.33 \\
5 & \multicolumn{1}{l|}{T (SUB1--SUB2+DESC2)}          & 0.413  &   0.372  &  3.555    &  0.09 & 0.09  &   0.08
  &  0.00    &  0.05  &  0.11  &  0.04  &  0.20  &  0.32 \\ 
6 &  \multicolumn{1}{l|}{T (SUB1+DESC1--SUB2)}          & 0.330  &   0.267  &  3.630    &  0.09 & 0.10  &   0.09
  &  0.00    &  0.26  &  0.12  &  0.04  &  0.23  &  0.35 \\ 
7 & \multicolumn{1}{l|}{T (SUB1+DESC1--DESC2)}          & 0.400  &   0.350  &  3.560    &  0.07 & 0.08  &   0.08
  &  0.00    &  0.00  &  0.10  &  0.03  &  0.19  &  0.35 \\ 
8 & \multicolumn{1}{l|}{T (SUB1+DESC1--SUB2+DESC2)}          & 0.417  &   0.378  &  3.530    &  0.05 & 0.07  &   0.08
  &  0.00    &  0.07  &  0.11  &  0.03  &  0.22  &  0.37 \\ 
9 &  \multicolumn{1}{l|}{T (SUB1+DESC1--SUB2+DESC2+SOL2)}          & 0.411  &   0.387  &  3.502    &  0.09 & 0.09  &   0.08
  &  0.00    &  0.06  &  0.12  &  0.04  &  0.20  &  0.40 \\
\hline
11 &  \multicolumn{1}{l|}{E (SUB1--SUB2)  ({\bf unsupervised baseline})}          & 0.141  &   0.108  &  3.636    &  0.39 & 0.38  &   0.36
  &  0.00    &  0.03  &  0.08  &  0.02  &  0.13  &  0.24 \\ 
12 &  \multicolumn{1}{l|}{E (DESC1--DESC2)}          & 0.034  &   0.059 &  4.201    &  {\bf 0.40} & {\bf 0.40}  &   {\bf 0.39}
  &  0.00    &  0.10  &  0.07  &  0.03  &  0.12  &  0.18 \\
13 &  \multicolumn{1}{l|}{E (SUB1+DESC1--SUB2+DESC2)}          & 0.103  &   0.051  &  5.210    &  0.16 & 0.16  &   0.15
  &  0.00    &  0.03  &  0.11  &  0.07  &  0.16  &  0.20 \\
14 &  \multicolumn{1}{l|}{E (SUB1+DESC1--SUB2+DESC2+SOL2)}          & 0.063  &   0.041  &  5.607    &  0.20 & 0.17  &   0.16
  &  0.00    &  0.03  &  0.13  &  0.05  &  0.13  &  0.22 \\
 \hline \hline
15 &  \multicolumn{1}{l|}{S-LSTM(SUB1--SUB2)  ({\bf supervised baseline})}          & 0.530  &   0.501  &  3.778    &  0.272 & 0.234  &   0.212
  &  0.000    &  0.128  &  0.080  &  0.022  &  0.111  &  0.311 \\ 
16 & \multicolumn{1}{l|}{S-LSTM (DESC1--DESC2)}           & 0.641  &   0.586  &  3.220    &  0.277 & 0.244  &   0.222
  &  0.100    &  {\bf 0.287}  &  0.209  &  0.111  &  0.3111  &  0.489  \\  
17 &  \multicolumn{1}{l|}{S-LSTM (SUB1+DESC1--SUB2+DESC2)}           & 0.662  &   0.621  &  2.992    &  0.288  & 0.251  &   0.232
&  0.137    &  0.129  &  0.208  &  0.111  &  0.342  &  0.511 \\   
18 &  \multicolumn{1}{l|}{S-LSTM (SUB1+DESC1--SUB2+DESC2+SOL2)}           & 0.693  &   0.631  &  2.908    &  0.298  & 0.236  &   0.241
&  0.143    &  0.189  &  0.228  &  0.133  &  0.353  &  0.548 \\   \hline
19 &  \multicolumn{1}{l|}{ML-LSTM (SUB1--SUB2, DESC1--DESC2)}      & 0.688  &   0.644  &  2.870    &  0.290 & 0.255  &   0.234
&  0.250    &  0.121  & 0.167  &  0.067  &  0.289  &  0.533 \\  
20 & \multicolumn{1}{l|}{ + CL-LSTM (SUB, DESC, SOL)}       & 0.744  &   0.680  &  2.470    &  0.293  & 0.259  &   0.238
&  0.143    &  0.179  &  0.286  &  {\bf 0.178}  &  0.378  &  0.564 \\       
21 &  \multicolumn{1}{l|}{ + weighted channels (h*0.8, E*0.2)}  & 0.758  &   0.701  &  2.354    &  0.392 & 0.376  &   0.346
&  {\bf 0.253}    &  0.176  & 0.248  &  0.111  &  0.439  &  0.579 \\
22 &  \multicolumn{1}{l|}{ + weighted channels (h*0.7, E*0.1, T*0.2)}  & {\bf 0.792}  &  {\bf  0.762}  &  {\bf 2.052}    &  0.382 & 0.356  &   0.344
&  0.242    &  0.202  & {\bf 0.288}  &  0.133  &  {\bf 0.493}  &  {\bf 0.618} 
\end{tabular}}
\caption{Results on Development set: Pearson correlation (\textit{r}), Spearman’s rank correlation coefficient ($\rho$), 
Mean Squared Error (MSE), Mean Average Precision@k (MAP@k), Mean Reciprocal Rank@k (MRR@k) and Accuracy@k (Acc@k) 
for the multi-level (ML) and cross-level (CL) similarity learning, and retrieving the k-most similar tickets for each query (SUB1+DESC1). 
{\bf \#[1-14]}: Unsupervised baselines with DocNADE (T) and SumEMB (E). {\bf \#[15-18]}: Supervised Standard Siamese baselines. {\bf \#[19-22]}: Supervised Replicated Siamese with multi-channel and cross-level features.}
\label{results}
\end{table*}

\section{Evaluation and Analysis}
We evaluate 
the proposed method on our industrial data for textual similarity learning and retrieval tasks in the ticketing system. 
Table \ref{modelsettings} shows the different model configurations used in 
the following experimental setups.
We use Pearson correlation, Spearman correlation and Mean Squared Error\footnote{http://alt.qcri.org/semeval2016/task1/} (MSE) metrics 
for STS and 9 different metrics (Table~\ref{results}) 
for IR task.

\subsection{Industrial Dataset for Ticketing System}
Our industrial dataset consist of queries and historical tickets. 
As shown in Table \ref{examplemotivation}, a query consists of {\it subject} and {\it description} texts, 
while a historical ticket in knowledge base (KB) consists of {\it subject}, {\it description} and {\it solution} texts.
The goal of the ITS is to automatically recommend an optimal action i.e. {\it solution} for an input query, retrieved from the existing KB. 

There are $\mathfrak{T} = 949$ historical tickets in the KB, out of which 421 pairs are labeled with their relatedness score. 
We randomly split the labeled pairs by 80-20\% for train ($P_{tr}$) and development ($P_{dev}$). 
The relatedness labels are: {\it YES} (similar that provides correct solution), 
{\it REL} (does not provide correct solution, but close to a solution) and {\it NO} (not related, not relevant  and provides no correct solution). 
 We convert the labels into
 numerical scores [1,5], where {\it YES}:5.0, {\it REL}:3.0 and {\it NO}:1.0. 
The average length (\#words) of  SUB, DESC and SOL are 4.6, 65.0 and 74.2, respectively. 

The end-user (customer) additionally supplies 28 unique queries ($Q_{U}$) (exclusive to the historical tickets)
to test system capabilities to retrieve the optimal solution(s) by computing $28\times949$ pairwise ticket similarities.  
We use these queries for the end-user qualitative evaluation for the $28\times10$ proposals (top 10 retrievals for each query). 



 
\def\arraystretch{1.25}
\begin{table*}[t]
\centering
\small
\resizebox{\textwidth}{!}{
\begin{tabular}{r|ccc|ccccccccc}
 \multicolumn{1}{c|}{\textbf{Model}} & \multicolumn{3}{c}{\textbf{Similarity Task}} & \multicolumn{9}{|c}{\textbf{Retrieval Task}} \\ \cline{2-13}
 & \multicolumn{1}{c}{\textit{r}} & \multicolumn{1}{c}{$\rho$} & \multicolumn{1}{c|}{MSE} & \multicolumn{1}{c}{MAP@1} & \multicolumn{1}{c}{MAP@5} & \multicolumn{1}{c}{MAP@10} & \multicolumn{1}{c}{MRR@1} & \multicolumn{1}{c}{MRR@5} & \multicolumn{1}{c}{MRR@10} & \multicolumn{1}{c}{Acc@1} & \multicolumn{1}{c}{Acc@5} & \multicolumn{1}{c}{Acc@10}\\ \hline


\multicolumn{1}{l|}{T (SUB1-SUB2)}          & 0.414  &   0.363  &  5.062    &  0.04 & 0.03  &   0.03
  &  {\bf 0.29}    &  0.24  &  0.10  &  0.01  &  0.17  &  0.28 \\ 
\multicolumn{1}{l|}{T (SUB1-DESC2)}          & 0.399  &   0.362  &  3.791    &  0.04 & 0.03  &   0.03
  &  0.00    &  0.05  &  0.07  &  0.03  &  0.12  &  0.19 \\ 
\multicolumn{1}{l|}{T (DESC1-SUB2)}          & 0.371  &   0.341  &  3.964    &  0.05 & 0.06  &   0.05
  &  0.25    &  0.07  &  0.11  &  0.04  &  0.21  &  0.33 \\
\multicolumn{1}{l|}{T (DESC1-DESC2)}          & 0.446  &   0.398  &  3.514    &  0.05 & 0.05  &   0.04
  &  0.00    &  0.04  &  0.10  &  0.04  &  0.18  &  0.34 \\
\multicolumn{1}{l|}{T (SUB1-SUB2+DESC2)}          & 0.410  &   0.370  &  3.633    &  0.05 & 0.04  &   0.04
  &  0.00    &  0.12  &  0.08  &  0.04  &  0.13  &  0.20 \\
\multicolumn{1}{l|}{T (SUB1+DESC2-SUB2)}          & 0.388  &   0.326  &  3.561    &  0.06  & 0.06  &   0.05
  &  0.25    &  {\bf 0.29}  &  0.13  &  0.05  &  0.22  &  0.38 \\ 
\multicolumn{1}{l|}{T (SUB1+DESC1-DESC2)}          & 0.443  &   0.396  &  3.477    &  0.04 & 0.04  &   0.04
  &  0.00    &  0.00  &  0.10  &  0.03  &  0.17  &  0.37 \\ 
\multicolumn{1}{l|}{T (SUB1+DESC1, SUB2+DESC2)}          & {\bf 0.466}  &   {\bf 0.417}  &  3.460    &  0.05 & 0.05  &   0.04
  &  0.00    &  0.06  &  0.11  &  0.03  &  {\bf 0.24}  &  0.37 \\ 
\multicolumn{1}{l|}{T (SUB1+DESC1, SUB2+DESC2+SOL2)}          & 0.418  &   0.358  &  {\bf 3.411}    &  {\bf 0.07} & 0.06  &   {\bf 0.06}
  &  0.00    &  0.09  &  {\bf 0.14}  &  0.05  &  0.20  &  {\bf 0.39} \\ \hline
\end{tabular}}
\caption{DocNADE ($M2$) performance for the queries $Q_{L}\in (P_{tr}+P_{dev})$ in the labeled pairs in unsupervised fashion.}
\label{resultsunsupervisedsuppl}
\end{table*}

\definecolor{YES}{HTML}{0BDA51}
\definecolor{REL}{HTML}{0000CD}
\definecolor{NO}{HTML}{E2062C}
\definecolor{UNK}{HTML}{987654}
\pgfplotsset{width=7.2cm, height=4.5cm,compat=1.5}
\begin{figure*}[t]
\begin{center}
\makebox[0.7\textwidth]{
\begin{tikzpicture}
\begin{axis}[
    ybar stacked,
	bar width=4pt,
    enlargelimits=0.02,
    legend style={at={(0.5,-0.20)},
      anchor=north,legend columns=-1},
    ylabel={\#Tickets (YES/REL/UNK)},
    symbolic x coords={
    q1, q2, q3, q4, q5, q6,
    q7, q8, q9, q10, q11, q12,
    q13, q14, q15, q16, q17
    },
    xtick=data,
    x tick label style={rotate=45,anchor=east},
    ymin=0,
    ymax=10,
   ymajorgrids = true,
    tick label style={font=\footnotesize},
    legend style={font=\footnotesize},
    label style={font=\footnotesize},
    enlarge x limits=0.03,
    legend style={at={(0.9,0.9)},anchor=north east}
    ]
\addplot+[ybar, fill=YES] plot coordinates {
  (q1,3) (q2,1) (q3,1) (q4,1) (q5,4) (q6,5)
  (q7,1) (q8,1) (q9,2) (q10,0) (q11,1) (q12,2)
  (q13,0) (q14,2) (q15,4) (q16,1) (q17,1)};
\addplot+[ybar, fill=REL] plot coordinates {
  (q1,1) (q2,1) (q3,0) (q4,0) (q5,1) (q6,0)
  (q7,2) (q8,0) (q9,0) (q10,2) (q11,0) (q12,0)
  (q13,2) (q14,0) (q15,0) (q16,0) (q17,1)};
\addplot+[ybar, fill=UNK] plot coordinates {
  (q1,0) (q2,0) (q3,0) (q4,0) (q5,0) (q6,0)
  (q7,0) (q8,0) (q9,0) (q10,0) (q11,0) (q12,0)
  (q13,0) (q14,0) (q15,0) (q16,0) (q17,0)};
\legend{\strut YES, \strut REL, \strut UNK}
\end{axis}
\end{tikzpicture}
\begin{tikzpicture}
\begin{axis}[
    ybar stacked,
	bar width=4pt,
    enlargelimits=0.02,
    legend style={at={(0.5,-0.20)},
      anchor=north,legend columns=-1},
    symbolic x coords={
    q1, q2, q3, q4, q5, q6,
    q7, q8, q9, q10, q11, q12,
    q13, q14, q15, q16, q17, q18,
    q19, q20, q21, q22, q23, q24
    },
    xtick=data,
    x tick label style={rotate=45,anchor=east},
    ymin=0,
    ymax=10,
    ymajorgrids = true,
    tick label style={font=\footnotesize},
    legend style={font=\footnotesize},
    label style={font=\footnotesize},
    enlarge x limits=0.03,
    legend style={at={(1,1)},anchor=north east}
    ]
\addplot+[ybar, fill=YES] plot coordinates {
  (q1,0) (q2,1) (q3,1) (q4,1) (q5,2) (q6,3)
  (q7,0) (q8,1) (q9,0) (q10,0) (q11,1) (q12,1)
  (q13,0) (q14,1) (q15,2) (q16,1) (q17,0)};
\addplot+[ybar, fill=REL] plot coordinates {
  (q1,0) (q2,0) (q3,0) (q4,0) (q5,0) (q6,0)
  (q7,1) (q8,0) (q9,0) (q10,1) (q11,0) (q12,0)
  (q13,0) (q14,0) (q15,0) (q16,0) (q17,0)
  };
\addplot+[ybar, fill=UNK] plot coordinates {
  (q1,10) (q2,9) (q3,9) (q4,9) (q5,8) (q6,7)
  (q7,10) (q8,9) (q9,10) (q10,9) (q11,9) (q12,9)
  (q13,10) (q14,9) (q15,8) (q16,9) (q17,10)
  };
\end{axis}
\end{tikzpicture}
\pgfplotsset{width=4.5cm,height=4.0cm,compat=1.5}
\begin{tikzpicture}
    \begin{axis}[
       title=Success Rate(\%),
        x tick label style={/pgf/number format/1000 sep=},
	x axis line style = { opacity = 0 }, 
 	axis y line = none,
        ybar,
        bar width=6pt,
	symbolic x coords = {3$>$=, 2$>$=, 1$>$=},
        xtick = data,
        tickwidth = 0pt, 
  	enlarge x limits = 0.2, 
        scaled x ticks = true,
        nodes near coords,
        legend style={at={(0.6,1.1)},anchor=north east},
	tick label style={font=\footnotesize},
    	legend style={font=\tiny},
    	label style={font=\footnotesize}    
    ]
        \addplot+[ybar,style={fill=REL}]
            coordinates {(3$>$=, 20.0) (2$>$=,25.0) (1$>$=,77)};

	\addplot+[ybar,style={fill=YES}]
            coordinates {(3$>$=, 25.0) (2$>$=,43) (1$>$=, 73)};

        \legend{YES+REL, YES}
    \end{axis}
\end{tikzpicture}
}
\end{center}
\caption{Evaluation on End-user Queries (sub-sample). 
UNK: Unknown. 
 (Left) Gold Data: The count of  similar (YES) and relevant (REL) tickets for each query (q1-q17). 
 (Middle) ITS Results: For each query, ITS proposes the top 10 
  YES/REL retrievals. 
The plot depicts 
  the count of YES/REL proposals matched out of the top 10 gold proposals for each q. 
UNK may include YES, REL or NO, not annotated in the gold pairs. 
(Right) Success Rate:
  YES: percentage of correct similar(YES) proposal out of the top 10; YES+REL: percentage of correct similar (YES) and relevant(REL)  
   proposals  out of the top 10.}
\label{end-user evaluation}
\end{figure*}
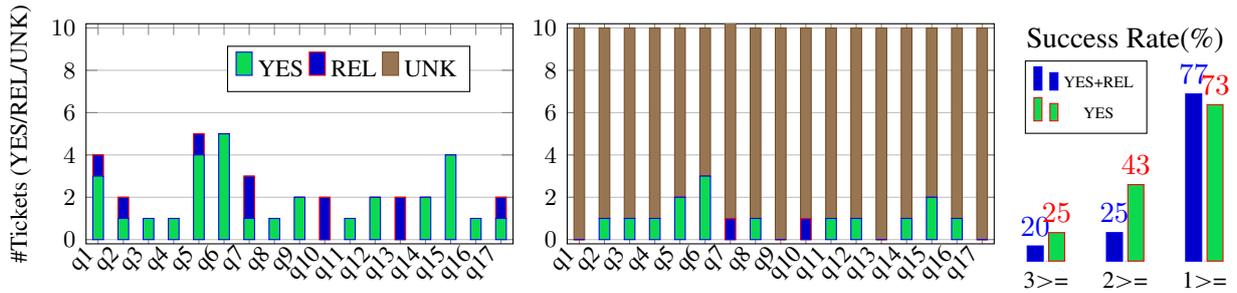
\subsection{Experimental Setup: Unsupervised}
We establish baseline for similarity and retrieval by the following two unsupervised approaches:

{\bf (1)} {\bf Topic Semantics} {\bf T}: 
As discussed in section \ref{DocNADE}, we use DocNADE topic model to learn document representation. 
To train, we take 50 held-out samples from the historical tickets $\mathfrak{T}$. 
We compute perplexity on 100 topics for each ticket component from the held-out set, 
comparing LDA and DocNADE models trained individually with SUB+DESC ($M1$) and SUB+DESC+SOL texts\footnote{+: merge texts to treat them as a single document} ($M2$).  
Table \ref{pplLDAvsDocNADE} shows that DocNADE outperforms LDA. 

Next, we need to determine which DocNADE model ($M1$ or $M2$) is less perplexed to the queries. 
Therefore, we use $M1$ and $M2$ to evaluate DESC1 and SUB1+DESC1 components of the two sets of queries: 
(1) $Q_{L}$  is the set of queries from labeled (421) pairs and (2) $Q_{U}$ is the end-user set. 
Table \ref{ppl} shows that $M2$ performs better than $M1$ for both the sets of queries with DESC1 or SUB1+DESC1 texts.
We choose $M2$ version of the DocNADE to setup baseline for the similarity learning and retrieval 
 in unsupervised fashion. 

To compute a similarity score for the given query $q$ and historical ticket $\mathfrak{t}$ where ($q$, $\mathfrak{t}$)$\in P_{dev}$, 
we first compute a latent topic vector ({\it T}) each for $q$ and $\mathfrak{t}$ using DocNADE ($M2$) 
and then apply the similarity metric $g$ (eq \ref{simmetric}). 
To evaluate retrieval for $q$, we retrieve the top 10 similar tickets, 
ranked by the similarity scores on their topic vectors.  
Table \ref{results} (\#No [1-9]) shows the performance of DocNADE for similarity and retrieval tasks. Observe that \#9 achieves  the best MSE ($3.502$)  and Acc@10 ($0.40$) out of [1-9], 
suggesting that the topic vectors of query (SUB1+DESC1) and historical ticket (SUB2+DESC2+SOL2) are the key in recommending a relevant SOL2.  
See the performance of DocNADE for all labeled pairs i.e. queries and historical tickets ($P_{tr}+P_{dev}$) in the Table \ref{resultsunsupervisedsuppl}.

{\bf (2)} {\bf Distributional Semantics} {\bf E}: 
Beyond topic models, we establish baseline using the SumEMB method (section \ref{replicatedSiamese}), where an  embedding vector {\it E} is computed  following the topic semantics approach.   
The experiments \#11-14 show that the SumEMB results in lower performance for both the tasks, suggesting a need of a supervised paradigm in order to learn similarities in asymmetric texts. 
Also, the comparison with DocNADE indicates that the topic features are important in the retrieval of tickets.

\subsection{Experimental Setup: Supervised}
For semantic relatedness scoring, we train the Replicated Siamese,  
using backpropagation-through-time under the Mean Squared Error
(MSE) loss function (after rescaling the
training-set relatedness labels to lie $\in$ [0, 1]). 
After training, we apply an additional non-parametric regression
step to obtain better-calibrated predictions $\in$ [1, 5], same as \cite{Mueller:81}. 
We then evaluate the trained 
model for IR task, where we retrieve the top 10 similar results (SUB2+DESC2+SOL2), ranked by their 
similarity scores,  for each query (SUB1+DESC1) in the development set and compute MAP@K, 
MRR@K and Acc@K, where K=1, 5, and 10. 

We use 300-dimensional pre-trained \textit{word2vec}\footnote{Publicly available at: code.google.com/p/word2vec}
 embeddings for input words, however, to generalize beyond the limited vocabulary in \textit{word2vec} 
due to industrial domain data with technical vocabulary, we also 
employ char-BLSTM \cite{Lample:81} to generate additional embeddings (=50 dimension\footnote{Run forward-backward character LSTM for every word and concatenate the last hidden units (25 dimension each)}).
  The resulting dimension for word embeddings is 350.  
 We use 50-dimensional hidden vector, $h_{t}$, memory cells, $c_{t}$
and Adadelta \cite{Zeiler:81} with dropout
 and gradient clipping \cite{Pascanu:81} for optimization. 
The topics vector ({\it T}) size is 100. 
We use python NLTK toolkit\footnote{http://www.nltk.org/api/nltk.tokenize.html} for sentence tokenization. 
See Table \ref{hyperparameterssupp} for the hyperparameters in Replicated Siamese LSTM  for experiment \#No:22. 

\subsection{Results: State-of-the-art Comparisons}\label{Results: STS and IR Tasks on Development Set}
Table~\ref{results} shows the similarity and retrieval scores for unsupervised and supervised baseline methods. 
The \#9, \#18 and \#20 show that the supervised approach performs better than unsupervised topic models. 
\#17 and \#19 suggest that the multi-level Siamese improves (Acc@10: 0.51 vs. 0.53) both STS and IR. Comparing \#18 and \#20, the cross-level Siamese shows performance gains (Acc@10: 0.55 vs. 0.57).
Finally, \#21 and \#22 demonstrates improved similarity (MSE: 2.354 vs. 2.052) and retrieval (Acc@10: 0.58 vs. 0.62) due to weighted multi-channel ($h$, $E$ and $T$) inputs.    


The replicated Siamese (\#22) with different features best results in 2.052 for MSE and 0.618 (= 61.8\%) for Acc@10.
We see 22\% and 7\% gain in Acc@10 for retrieval task, respectively over unsupervised (\#9 vs. \#22: 0.40 vs. 0.62) and supervised 
(\#18 vs. \#22: 0.55 vs. 0.62) baselines. 
The experimental results suggest that the similarity learning in supervised fashion improves the ranking of relevant tickets. 

\begin{table*}[t]
\centering
\resizebox{0.99\textwidth}{!}{
\begin{tabular}{c||c|c|c}
{\bf Query}                      & {\bf Recommendation\_1}                        &  {\bf Recommendation\_2}                        & {\bf Recommendation\_3}                   \\ \hline
\begin{tabular}[c]{@{}l@{}}
{\bf SUB:} \\
\underline{GT} \underline{Trip} - \underline{Low} \underline{Frequency} \underline{Pulsations}\\ \\
{\bf DESC:} \\
\underline{GT} \underline{Tripped} due to a sudden\\
 increase in \underline{Low} \underline{Frequency}\\
 \underline{Pulsations}. The \underline{machine} has been\\
 restarted and is now operating\\
 normally. \underline{Alarm} received was: \\
\underline{GT} XXX \underline{Low} \underline{Frequency} \underline{Pulsation}
\end{tabular} 
& \begin{tabular}[c]{@{}l@{}}
{\bf SUB:}\\
Narrow \underline{Frequency} \underline{Pulsations}\\  \\
{\bf DESC:}\\
\underline{Low} and Narrow \underline{frequency} \\
\underline{pulsations} were detected. \\
The peak value for the \underline{Low} \\
\underline{Frequency} \underline{Pulsations} is \\
\#\# mbar. \\ \\
{\bf SOL:} \\
XXXXX combustion support is \\
currently working on the issue. \\
The recommended action for \\
now is that the \underline{machine} XXXX \\
at {\bf {\bf load}} XXXX \#\# {\bf MW}. \\
\end{tabular}
& \begin{tabular}[c]{@{}l@{}}
{\bf SUB:} \\
\underline{Low} \underline{frequency} \underline{pulsations}\\ \\ 
{\bf DESC:}\\
High level \underline{low} \underline{frequency}\\ 
\underline{pulsations} were detected \\
when active {\bf load} is XXXX.\\ \\ 
{\bf SOL:}\\
Since the \underline{machine} is running\\
 with XXXX, the XXX  be \\
changed in the register.
 After\\
adjustment  is complete, monitor \\
 the \underline{machine} behavior between \\
\#\# {\bf MW} to \#\#  {\bf load}. 
\end{tabular}
& \begin{tabular}[c]{@{}l@{}}
{\bf SUB:} \\ 
\underline{GT3} - High \underline{Low} \underline{Frequency} \underline{Pulsation} alarms after \underline{trip}\\ 
{\bf DESC:} \\ 
Yesterday, after Steam Turbine \underline{tripped}, \underline{GT}-3\\ 
experienced high \underline{Low} \underline{Frequency} \underline{Pulsation} \underline{alarm}.\\
The{\bf load} of \underline{GT}-3 was \#\# {\bf MW} and\\
 went up as high as \#\#  {\bf MW}. During the time, \\
\underline{Low} \underline{Frequency} \underline{Pulsation} for 3 
 \underline{pulsation} \\devices went up as high as \#\#. 
The \underline{Low} \\ 
\underline{frequency} \underline{pulsation} was a XXX.\\
{\bf SOL:} \\
A {\bf load} XXXX from \#\# {\bf MW} to \#\# {\bf MW} is an\\
event XXX the unit 
XXXX \underline{trip}. The XXXX 
to \\
\underline{low} \underline{frequency} \underline{pulsation} during similar event,\\ should be XXXX. 
Check  that XXXX from after \\
the XXXX (XX005/XX01) into combustion chamber\\
 (XX030/XX01), XXXX should be XXXX. Repeat \\
 until XXXX is within the range of \#\# -\#\#.
\end{tabular}\\ \hline
{\bf (Rank, Similarity Score)}                  & {(1, 4.75)}                        & {(2, 4.71)}                         & {(3, 4.60)}  \\ \hline
{\bf \#Topics} \{{\bf \#83}, \#7, \#30\}                      & \{{\bf \#83}, \#16, \#30\}                        & \{\#7, {\bf \#83}, \#19\}                         & \{\#7, {\bf \#83}, \#19\} 
\end{tabular}}
\caption{Top-3 Tickets Retrieved and ordered by their (rank, similarity score) for an input test query. $\#Topics$: the top 3 most probable associated topics. 
{\bf SOL} of the retrieved tickets is returned as recommended action.  \underline{Underline}: Overlapping words; XXXX and \#\#: Confidential text and numerical terms.
}
\label{examplequalitative}
\end{table*}
\subsection{Success Rate: End-User Evaluation}
We use the trained similarity model to retrieve 
 the top 10 similar tickets from KB for each end-user query $Q_{U}$, 
and compute the number of correct similar 
 and relevant tickets. 
 For ticket ID $q6$ (Figure~\ref{end-user evaluation}, Middle), 3 out of 10 proposed tickets are 
 marked similar, where the end-user expects 4 similar tickets (Figure~\ref{end-user evaluation}, Left). 
 For ticket ID $q1$, $q13$ and $q17$,  
 the top 10 results do not include the corresponding expected tickets due to no term matches  
 and we find that the similarity scores for all the top 10 tickets 
 are close to 4.0 or higher, which indicates that the system proposes more similar tickets (than the expected
 tickets), not included in the gold annotations. 
  The top 10 proposals are evaluated for each query 
 by success rate 
(success, if N/10 proposals supply the expected solution).  
 We compute success rate (Figure~\ref{end-user evaluation}, Right) for (1 or more), (2 or more)
  and (3 or more) 
  correct results out of the top 10 proposals. 

\section{Qualitative Inspections for STS and IR}
Table \ref{examplequalitative} shows a real example for an input query, where the top 3 recommendations are proposed from the historical tickets using the trained Replicated Siamese model.  The recommendations are ranked by their similarity scores with the query.  The underline shows the overlapping texts. 

We also show the most probable topics (\#) that the query or each recommendation is associated with. 
The topics shown (Table \ref{exampletopics}) are learned from DocNADE model and are used in multi-channel network. 
Observe that the improved retrieval scores (Table \ref{results} \#22) are attributed to the overlapping topic semantics in query and the top retrievals. For instance, the
topic \#83 is the most probable topic feature for the query and recommendations. 
We found terms, especially {\it load} and {\it MW} in SOL (frequently appeared for other {\it Frequency Pulsations} tickets) that are captured in topics \#7 and \#83, respectively.

\section{Related Work}

Semantic Textual Similarity has diverse applications in 
information retrieval \cite{Larochelle:82,GuptaiDocNADE:82}, search, summarization \cite{Gupta:81}, recommendation systems, etc.  
For shared STS task in SemEval 2014, numerous researchers applied competitive methods 
that utilized both heterogeneous features (e.g. word overlap/similarity, negation
modeling, sentence/phrase composition) as well as external
resources (e.g. Wordnet \cite{Miller:81}), along with machine learning approaches such as
LSA \cite{Lan:81} and  word2vec neural language model \cite{Mikolov:81}.    
In the domain of question retrieval \cite{LiCai:82,KaiZhang:82}, users retrieve   
historical questions which precisely match their
questions (single sentence) semantically equivalent or relevant. 

Neural network based architectures, especially CNN \cite{Yin:81}, LSTM \cite{Mueller:81}, RNN encoder-decoder \cite{Kiros:81}, etc.
have shown success in similarity learning task in Siamese framework \cite{Mueller:81,Chopra:81}.  
These models are adapted to similarity learning in sentence pairs using complex learners. \newcite{Wieting:81}  observed that
 word vector averaging and LSTM for similarity learning perform better in short and long text pairs, respectively. Our learning objective exploits the multi-channel representations of short and longer texts and compute cross-level similarities in different components of the query and tickets pairs.  
Instead of learning similarity in a single sentence pair, we propose a novel task 
and neural architecture for asymmetric textual similarities.  
To our knowledge, 
this is the first advancement of Siamese architecture towards multi-and-cross level similarity 
learning in asymmetric text pairs with an industrial application.

\begin{table}[t]
\centering
\resizebox{0.8\textwidth}{!}{
\begin{tabular}{r|l}
{\bf ID} & \multicolumn{1}{c}{\bf Topic Words (Top 10)}\\ \hline
\#83 &  pulsation, frequency, low, load, high, pulsations, increase, narrow, XXXX, mw\\
\#7 &  trip, turbine, vibration, gas, alarm, gt, time, tripped, pressure, load\\
\#30 & start, flame, unit, turbine, combustion, steam, temperature, compressor, XXXX, detector\\
\#16 & oil, XXXX, XXXX, pressure, kpa, dp, level, high, mbar, alarm\\
\#19 &  valve, XXXX, fuel, valves, gas, bypass, check, control, XXXX, XXXX
\end{tabular}}
\caption{Topics Identifier and words captured by DocNADE}
\label{exampletopics}
\end{table}

\section{Conclusion and Discussion}
We have demonstrated 
deep learning application in STS and 
IR tasks for an industrial ticketing system. 
The results indicate that the proposed LSTM is capable of modeling complex semantics 
by explicit guided representations and does not rely on hand-crafted linguistic features, therefore being generally applicable 
to any domain. 
We have showed improved similarity and retrieval via the proposed multi-and-cross-level Replicated Siamese architecture, leading to relevant recommendations especially in industrial use-case. 
As far we we know, 
this is the first advancement of Siamese architecture for similarity 
learning and retrieval in asymmetric text pairs with an industrial application.

We address the challenges in a real-world industrial application of ticketing system. 
Industrial assets like power plants, production lines, turbines, etc. need to be serviced well because an unplanned outage always leads to significant financial loss. It is an established process in industry to report issues (via query) i.e. symptoms which hint at an operational anomaly to the service provider. This reporting usually leads to textual descriptions of the issue in a ticketing system. The issue is then investigated by service experts who evaluate recommended actions or solutions to the reported issue. The recommended actions or solutions are usually attached to the reported issues and form a valuable knowledge base on how to resolve issues. Since industrial assets tend to be similar over the various installations and since they don't change quickly it is expected that the issues occurring over the various installations may be recurring. Therefore, if for a new issue similar old issues could be easily found this would enable service experts to speed up the evaluation of recommended actions or solutions to the reported issue. The chosen approach is to evaluate the pairwise semantic similarity of the issues describing texts.

We have compared unsupervised and supervised approach for both similarity learning and retrieval tasks, where the supervised approach leads the other. 
However, we foresee significant gains with the larger amount of similarity data as the amount of labeled similarity data grows and the continuous feedback is incorporated for optimization within the industrial domain, where quality results are desired. 
In future work, we would also like to investigate attention \cite{bahdanau+al-2014-nmt} mechanism and dependency \cite{socher2012semantic,GuptacrossRE:82} structures in computing tickets' representation. 


\section*{Acknowledgements}

We thank our colleagues  Mark Buckley,  Stefan Langer, Subburam Rajaram and Ulli Waltinger, and 
anonymous reviewers for their review comments. 
This research was supported by Bundeswirtschaftsministerium ({\tt bmwi.de}), grant 01MD15010A (Smart Data Web) 
at Siemens AG- CT Machine Intelligence, Munich Germany.

\bibliographystyle{acl}
\bibliography{coling2018}

\end{document}